\documentclass[10pt,a4paper]{article}
\usepackage[T2A]{fontenc}
\usepackage[utf8]{inputenc}
\usepackage[english]{babel}
\usepackage{amssymb,graphicx}
\usepackage{amsmath,amsfonts}
\usepackage{amsthm}
\usepackage{framed, float}
\usepackage{feynmf}
\usepackage{microtype}
\usepackage{hyperref}

 \newcommand{\tr}{\mathop{\mathrm{tr}}\nolimits}

\hoffset= - 1.0in         % switch off for draft style
\title{{\Large \textbf{Refined toric branes, surface operators and
      factorization\\
      of generalized Macdonald
      polynomials}}\vspace{0.2cm}} \date{} \author{
  \textbf{Yegor Zenkevich}\footnote{yegor.zenkevich@gmail.com}\vspace{0.2cm}\\
  {\small {\it ITEP, Moscow 117218, Russia}}\\
  {\small {\it National Research Nuclear University MEPhI, Moscow
      115409, Russia }}\\
  {\small {\it Physics Department, Moscow State University, Moscow
      117312, Russia
    }}\\
  {\small {\it Dipartimento di Fisica, Universit\`a di Milano-Bicocca,
      Piazza della Scienza 3, I-20126 Milano, Italy}}\\
  {\small {\it INFN, sezione di Milano-Bicocca, I-20126 Milano, Italy
    }} }
\begin{document}
\maketitle
\vspace{-8.5cm}

\begin{center}
  \hfill ITEP/TH-37/16
\end{center}

\vspace{7.5cm}

\begin{abstract}
  We find new universal factorization identities for generalized
  Macdonald polynomials on the topological locus. We prove the
  identities (which include all previously known forumlas of this
  kind) using factorization identities for matrix model averages,
  which are themselves consequences of Ding-Iohara-Miki
  constraints. Factorized expressions for generalized Macdonald
  polynomials are identified with refined topological string
  amplitudes containing a toric brane on an intermediate preferred
  leg, surface operators in gauge theory and certain degenerate CFT
  vertex operators.
\end{abstract}

\section{Introduction.}
\label{sec:introduction}
The interplay between algebraic structures and geometry has been
fundamental to the development of mathematics in the recent
decades. In particular, it has led to a cornucopia of new results in
mathematical physics. One of examples is the (refined) topological
vertex function~\cite{top-vert, ref-top} which on the geometric side
describes GW and DT invariants of toric Calabi-Yau (CY) three-folds,
while from the algebraic point of view it is the intertwiner of the
Ding-Iohara-Miki (DIM) algebra\footnote{Alternative names include
  quantum toroidal, elliptic Hall, spherical degenerate DAHA algebras,
  or simply
  $U_{q,t}(\widehat{\widehat{\mathfrak{gl}}}_1)$.}~\cite{DIM}. The
second famous example comes from the gauge theory: the equivariant
cohomology of the instanton moduli spaces (captured by Nakajima quiver
varieties~\cite{Nakajima} and the corresponding Nekrasov partition
functions~\cite{Nekr}) is acted on by a certain vertex operator
algebra, which turns out to be the $W_N$-algebra of two dimensional
conformal field theory. This correspondence between the geometric
(moduli space) and algebraic ($W_N$-algebra) objects is known as the
AGT relation~\cite{AGT} and has many known implications and
generalizations~\cite{AGTmore}. These two examples are in fact
directly related to each other and their relation can be understood on
both sides of the algebro-geometric correspondence. On the algebraic
side the equivariant cohomology (or more precisely $K$-theory) of the
instanton moduli spaces is a tensor product of Fock representations of
the DIM algebra~\cite{AFS}, while the $q$-deformed $W_N$-algebra
generators are built from the currents of the DIM
algebra~\cite{AFHKSY, MMZ, AKTMMMOZ} and vertex operators are
combinations of topological vertices intertwining the action of the
DIM algebra and therefore of the $W_N$-algebra~\cite{Matsuo}. On the geometric side the $5d$ gauge theory is obtained
by compactifying M-theory on the toric CY three-fold. The parameters
of the gauge theory correspond to K\"ahler moduli of the CY and the
cohomology of the moduli space of instantons is identified with the
Hilbert space of M2 branes stretching between toric fixed points.

In this paper we explore a particular case of the algebro-geometric
correspondence, which is important for topological strings as well as
for gauge theories. We consider refined toric branes wrapping
Lagnangian submanifolds inside a toric CY
three-fold~\cite{AgSh-brane}. As is well-known, this setup corresponds
to surface operators in gauge theory and to degenerate fields of the
$W_N$-algebra~\cite{Aganagic-surf, Pasq, DGH, Taki}. We will consider
mostly the algebraic side of the problem and relate the stack of
refined branes on the preferred leg of the toric diagram to a
particular intertwining operator of DIM algebra, which can be recast
into a combination of generalized Macdonald polynomials~\cite{O,
  Z}. The properties of the branes are related to the remarkable
\emph{factorization} identities for generalized polynomials evaluated
on a particular submanifold in the brane moduli space called the
\emph{topological locus.}

Generalized Macdonald polynomials~\cite{O, Z} play the central role
in the AGT correspondence. They arise naturally in the study of the
DIM algebra representations on tensor producs of Fock
modules. In~\cite{Z},~\cite{MZ} matrix elements between generalized
Macdonald polynomials were computed using matrix model
techniques. They turned out to reduce to \emph{integral factorization
  identities,} which provide a very explicit answer for the
$q$-Selberg averages in terms of Nekrasov functions.

In this note we would like to use these integral identities to prove
new \emph{topological locus} factorization identity recently found
in~\cite{KM}. For special values of parameters the integrals disappear
and one is left with Macdonald polynomials evaluated at the
topological locus. The integral identity implies that those are still
given by the factorized formulas. This technique allows us to find
several new identities for generalized Macdonald polynomials on a more
general topological locus. In this way we prove and generalize the
results of~\cite{KM}.

We then connect the factorization of the polynomials to the refined
topological string picture. To this end we will interpret matrix model
averages as topological string amplitudes on toric CY threefolds with
Lagrangian branes appropriately placed on the legs of the toric
diagram (for exact correspondence and explanation
see~\cite{MZ}). Factorization of averages relies on the particular
properties of the branes residing on the preferred direction of the
diagram. The topological locus corresponds to a certain degenerate
limit of the CY, which models addition of a stack toric branes on one
of the legs in the preferred direction. Factorization of generalized
Macdonald polynomials in this picture allows us to understand the
amplitudes with toric branes placed on intermediate preferred legs of
the toric diagram.

One can also interpret factorization formulas for generalized
Macdonald polynomials in terms of CFT vertex operators in
Dotsenko-Fateev (DF) representation. In this case the radical
simplification of the formulas occurs due to the particular choice of
the dimensions for which the vertex operators do not require any
screening currents. In view of the AGT relations this corresponds to a
particularly simple surface operator in the corresponding gauge
theory.

In the remaining part of the introduction we discuss the main points
of this network of correspondences in more detail. In
sec.~\ref{sec:fact-gener-macd} we write down and prove the
factorization identities, in sec.~\ref{sec:expl-from-refin} we connect
this results with topological strings and gauge theories. We present
our conclusions in sec.~\ref{sec:concl-furth-prosp}.

\subsection{Refined topological strings and branes}
\label{sec:refin-topol-strings}
Refined topological string theory~\cite{ref-top} is a deformation of
the topological string theory living on a toric CY three-fold, which
gives additional information on the spin content of the D-brane BPS
spectrum of type IIA string theory. Refined amplitudes are computed
using refined topological vertex, quite similarly to the ordinary
topological vertex computations~\cite{top-vert}. A Young diagram $Y_i$
is assigned to each leg $i$. The vertices, always trivalent,
correspond to certain explicit combinations $C_{Y_i Y_j Y_k}(q,t)$ of
symmetric functions depending on three Young diagrams~$Y_{i,j,k}$
on the adjacent legs~$i$, $j$, $k$:
\begin{equation}
  \label{eq:24}
  \parbox{2.4cm}{\includegraphics[width=2.4cm]{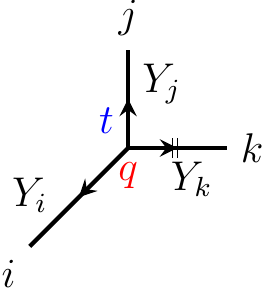}}\quad
  = \quad C_{Y_i Y_j Y_k}(q,t)
\end{equation}
There is one crucial point in the computation of \emph{refined}
amplitudes. Unrefined topological vertex $C_{Y_i Y_j Y_k} (q,q)$ is
cyclically symmetric in the three Young diagrams $Y_i$, $Y_j$, $Y_k$,
while the refined vertex $C_{Y_i Y_j Y_k} (q,t)$ for general $q\neq t$
is not. The recipe above, therefore, includes the choice of ordering
of Young diagrams in each vertex. This choice is indicated by the
double ticks and the labels $q$ and $t$ on the corresponding legs in
Eq.~\eqref{eq:24}. In what follows we will usually omit the indices
$q$ and $t$. It turns out that the choices for the neighbouring
vertices should be coordinated, so that the only freedom remaining is
the \emph{global} choice of the \emph{preferred direction} (horizontal
in Eq.~\eqref{eq:24}) on the toric diagram. We omit here the concrete
expression for $C_{Y_i Y_j Y_k}(q,t)$, which can be easily found in
the literature, not to overcomplicate our presentation.

To get the final answer for the amplitude one takes the sum over all
the Young diagrams $Y_i$ on the intermediate legs, each taken with
weight\footnote{In general there are also framing factors which we
  will not need here.}  $(-Q_i)^{|Y_i|}$, where $Q_i$ denotes the
exponentiated complexified K\"ahler parameter of the two-cycle
associated to the leg $i$. Let us give the simplest example of two
vertices glued together to form the resolved conifold geometry:
\begin{equation}
  \label{eq:25}
  \parbox{4cm}{\includegraphics[width=4cm]{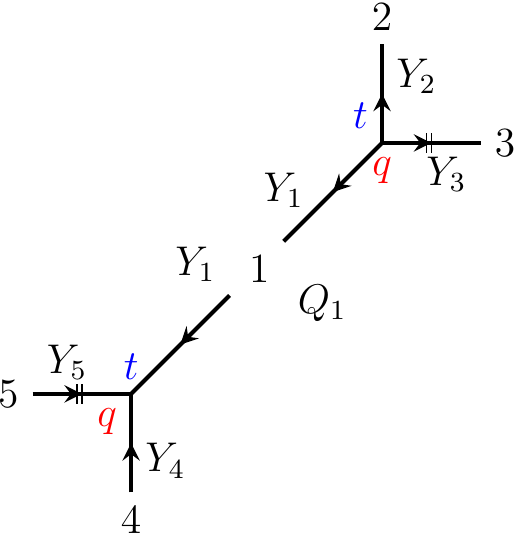}}\quad
  = \quad \sum_{Y_1} (-Q_1)^{|Y_1|} C_{Y_1 Y_2 Y_3}(q,t) C_{Y_1^{\mathrm{T}} Y_4^{\mathrm{T}} Y_5^{\mathrm{T}}}(t,q) 
\end{equation}

With the external lines one can associate either empty or non-empty
diagrams which do not take part in the sums. The former choice gives
the \emph{closed} string amplitude (partition function), while the
latter one gives the \emph{open} string amplitude with stacks of toric
branes on the external legs determining the external diagrams, or
``boundary conditions'' for the theory:
\begin{gather}
  \label{eq:26}
  \parbox{2.8cm}{\includegraphics[width=2.8cm]{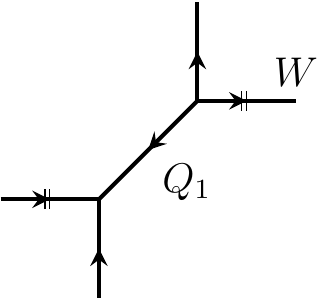}}= \parbox{3.6cm}{\includegraphics[width=3.6cm]{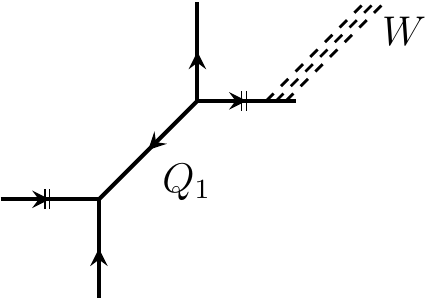}}
  \quad = \quad \sum_{Y_1} (-Q)^{|Y_1|} C_{Y_1 \varnothing W}(q,t)
  C_{Y_1 \varnothing \varnothing}(t,q) = Z_{\mathrm{open}}\left(
    \begin{smallmatrix}
      & \varnothing & \\
      \varnothing & & W\\
      & \varnothing &
    \end{smallmatrix}
\Big|Q_1\right)\\
  Z_{\mathrm{closed}}(Q_1) = Z_{\mathrm{open}}\left(
    \begin{smallmatrix}
      & \varnothing & \\
      \varnothing & & \varnothing\\
      & \varnothing &
    \end{smallmatrix}
\Big|Q_1\right)
\end{gather}
The dashed lines here denote toric branes. The number of branes in the
stack sets the maximal possible number of rows in the Young diagram
$W$. The final answer for the closed string amplitude does not depend
on the choice of the preferred direction, though open string
amplitudes do\footnote{In the algebraic approach of~\cite{AFS} the
  choice of the preferred direction is associated with the choice of
  the \emph{slope} of the coproduct $\Delta$ used in the definition of
  the DIM algebra. The most relevant choices used e.g.\ in~\cite{FJMM}
  where the ``horizontal'' coproduct $\Delta$ and the ``vertical'' (or
  perpendicular, or Drinfeld) coproduct $\Delta^{\perp}$.}.

In the unrefined case there is also a natural way to put a stack of
toric branes on the \emph{internal} leg (and indeed on any Lagrangian
submanifold of the CY). However, in the refined case only branes on
the \emph{external} lines have been considered so far. In the present
paper we will address this problem and propose a way to put a stack of
branes on the intermediate preferred leg. To do this we will employ
the duality between open and closed string amplitudes.

Open-closed duality in topological strings allows one to model stacks
of toric branes by \emph{closed} string amplitudes~\cite{Pasq, DGH,
  Taki}. The open string amplitudes should be packed in the
Ooguri-Vafa generating function, and the closed strings propagate in
the \emph{modified background} containing additional vertical line in
the toric diagram. Let us draw the dual pictures in the simplest case
of \emph{one} toric brane. The diagram corresponding to the brane can
have at most one column, i.e.\ it is of the form\footnote{Compared to
  the notation of~\cite{Taki} we use the transposed diagram
  $W$. Another way to obtain our conventions from that of~\cite{Taki}
  is to exchange the equivariant parameters, $q \leftrightarrow
  t^{-1}$. Using the terminology of~\cite{AgSh-brane} this amounts to
  the exchange of $q$-branes and $\bar{t}$-branes.} $W = [l]$. We then
have
\begin{equation}
  \label{eq:27}
 \sum_{l\geq 0} z^l  \parbox{3.6cm}{\includegraphics[width=3.6cm]{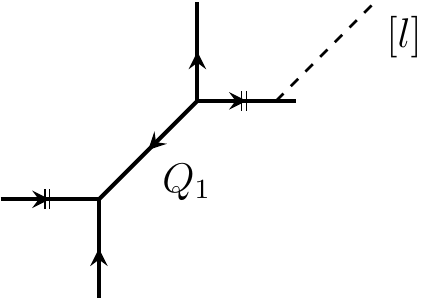}}
   =  \parbox{4cm}{\includegraphics[width=4cm]{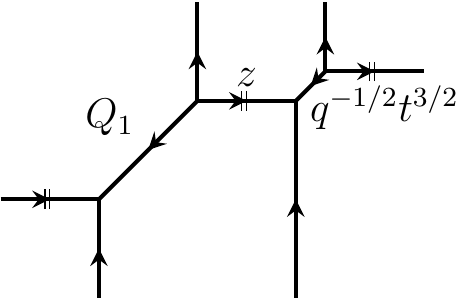}}
\end{equation}
In the l.h.s.\ of Eq.~\eqref{eq:27} $z$ plays the role of the holonomy
of the (abelian) gauge field living on the toric brane, while on the
r.h.s.\ it is identified with the K\"ahler parameter of the two-cycle
obtained by adding an extra vertical line to the geometry. In general,
for $N$ toric branes the K\"ahler parameter in the r.h.s.\ will change
to $q^{-1/2} t^{N + 1/2}$.

There are several points requiring clarification in this approach
which are absent for ordinary topological string, i.e.\ for $t=q$, and
appear only in the refined case:
\begin{enumerate}
\item The additional vertical line necessarily intersects all the
  parallel legs coming out of the diagram if there happen to be any
  (see Fig.~\ref{fig:6} b) for an example). One expects that the
  amplitude should be insensitive to these intersections since they
  have nothing to do with the toric brane insertion. However, in the
  refined case there is no way to make a ``trivial crossing'' of two
  lines: no choice of the K\"ahler parameter gives the desired
  result. One concludes that for several parallel legs the toric brane
  attached to one of them also interacts with all the others.

\item Although there is no way to make a ``trivial crossing'' of lines
  one can make a crossing, which models the trivial one in some
  situations. For example, this crossing can be used to set the
  diagram on one side to vanish if the diagram on the other vanishes
  (see Fig.~\ref{fig:6} b), c)). However, it works only in \emph{one}
  direction: either the left diagram vanishes whenever the right one
  is empty or vice versa.

\item Because of these features of the refined theory it is unclear
  how to put a toric brane on the intermediate preferred leg.
\end{enumerate}
The explanation of these puzzles will be the main focus of the present
work. We will show that the amplitudes in the presence of the toric
brane on the intermediate leg can be identified with generalized
Macdonald polynomials evaluated on the \emph{topological locus.}

\subsection{$q$-deformed CFT}
\label{sec:q-deformed-cft}

\begin{figure}[h]
  \centering
  \includegraphics[width=14cm]{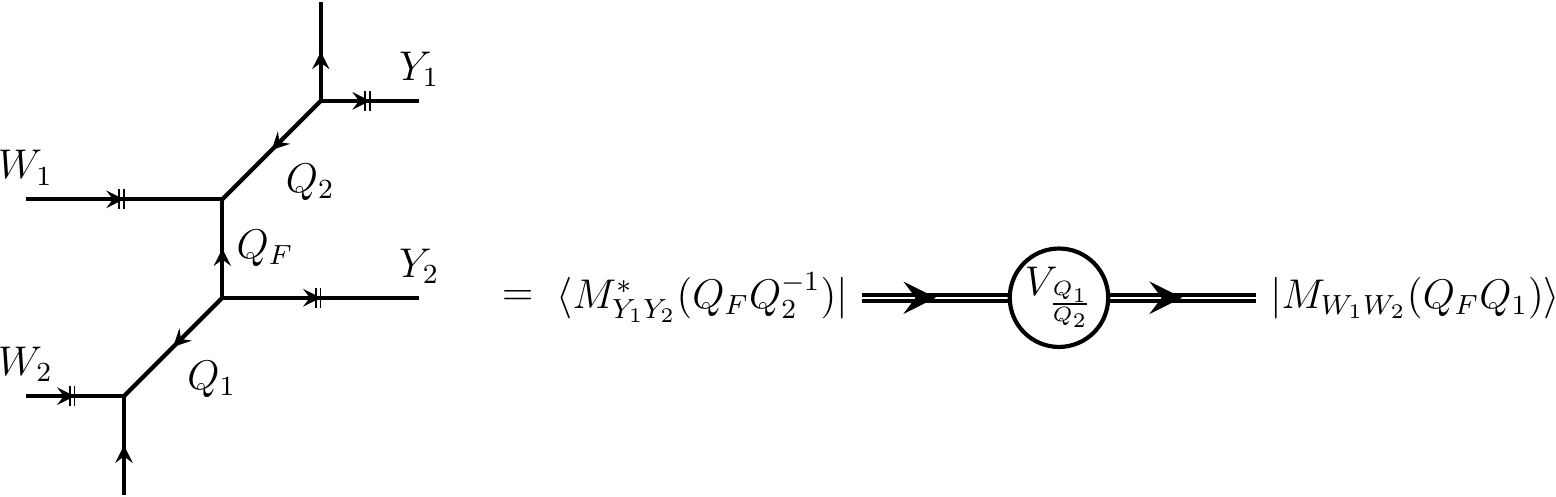}
  \caption{The correspondence between refined topological string
    amplitudes and vertex operators of CFT. Double lines in the
    r.h.s.\ denote the CFT Hilbert space on which $\mathsf{Vir}_{q,t}
    \oplus \mathsf{Heis}$ algebra acts. On the l.h.s.\ it corresponds
    to the two horizontal lines. The circle in the r.h.s.\ represents
    the vertex operator corresponding to the intersection with the
    vertical line in the toric diagram in the l.h.s. The matrix
    element on the r.h.s.\ is computed in the basis of generalized
    Macdonald polynomials $M_{Y_1 Y_2}$, which corresponds to the
    choice of horizontal preferred direction (lines marked by by
    double ticks) on the l.h.s. Notice the relation between the
    K\"ahler parameters $Q_{1,2}$, $Q_F$ of the CY on the l.h.s.\ and
    the parameters of the vertex operator $V_{Q_1/Q_2}$ and the states
    on the r.h.s.}
  \label{fig:10}
\end{figure}

It was shown in~\cite{Z,MZ}, that certain refined topological string
amplitudes on toric CY three-folds correspond to conformal blocks of
the $q$-deformed Virasoro or $W_N$-algebras\footnote{More concretely,
  to get a conformal block one should consider only \emph{balanced}
  toric diagrams, see~\cite{MMZ,AKTMMMOZ} for details.}. The horizontal
legs of the toric diagram represent the Hilbert space of the CFT, on
which the conformal algebra acts, and the intersections with vertical
legs give vertex operators or intertwiners of the algebra (see
Fig.~\ref{fig:10}). Naturally, the sums over Young diagrams living on
the \emph{horizontal} lines represent the sums over the complete basis
of states in the CFT Hilbert space. There is a natural choice for such
a basis --- the basis of generalized Macdonald polynomials, which
leads to explicit \emph{factorized} matrix elements for the vertex
operators given by Nekrasov formulas. The sums over diagrams on the
\emph{vertical} lines corresponds to the integrals over the positions
of the \emph{screening currents} appearing in the Dotsenko-Fateev
representation of the conformal blocks. Therefore, vertical lines
correspond not simply to vertex operators, but more concretely to the
\emph{screened} vertex operator insertions~\cite{MZ,AKTMMMOZ}. The
topological loci, i.e.\ the submanifold of the moduli space on which
generalized Macdonald polynomials \emph{factorize} into products of
monomials, represent the special set of parameters, for which the
number of screenings is zero.

\section{Factorization of generalized Macdonald polynomials}
\label{sec:fact-gener-macd}

\subsection{Schur and Macdonald polynomial factorization. A reminder}
\label{sec:schur-macd-polyn}
Let us first recall the familiar factorization formulas Schur and
Macdonald polynomials. Schur polynomials $s_Y(x_i)$ are symmetric
polynomials in the variables $x_i$, $i=1\ldots N$ labelled by Young
diagrams $Y$. They can be understood as characters of
finite-dimensional irreducible representations $\mathcal{R}_Y$ of
$\mathfrak{sl}_N$ algebra corresponding to the Young diagrams $Y$:
\begin{equation}
  \label{eq:31}
  s_Y(x) = \tr_{\mathcal{R}_Y} \mathrm{diag} (x_1, \ldots , x_N)
\end{equation}
We usually write all symmetric polynomials as functions of the power
sums $p_n = \sum_{i=1}^N x_i^n$. For particular values of the
variables lying on the \emph{topological locus} $p_n = \frac{1 -
  A^n}{1-q^n}$ Schur polynomials are given by very simple
\emph{factorized} formulas:
\begin{equation}
  \label{eq:30}
  s_Y \left( \frac{1 - A^n}{1-t^n} \right) = \prod_{(i,j) \in Y}
  t^{i-1} \frac{1 - A t^{j-i}}{1 - t^{Y_i-j + Y^{\mathrm{T}}_j - i + 1} } 
\end{equation}
These expressions can be related to ``quantum dimensions'', or
generating functions of the values of the Casimir operators on the
corresponding representations.

Macdonald polynomials $M_Y^{(q,t)}(p_n)$ provide a natural
generalization of Schur polynomials, depending on two parameters $q$
and $t$. Macdonald polynomials do not have immediate group theory
interpretation, but nevertheless have many properties similar to Schur
polynomials, to which they reduce for $t=q$. In particular, they still
factorize on the topological locus $p_n = \frac{1 - A^n}{1 - t^n}$:
\begin{equation}
  \label{eq:33}
  M_Y^{(q,t)} \left( \frac{1 - A^n}{1-t^n} \right) = \prod_{(i,j) \in Y}
  t^{i-1} \frac{1 - A q^{j-1} t^{1-i}}{1 - q^{Y_i-j} t^{Y^{\mathrm{T}}_j - i + 1} } 
\end{equation}
Notice that the parameters of the topological locus for Macdonald
polynomials are tied with the deformation parameters, so that for
given $t$ the locus is one-dimensional. We will see similar effect in
the following sections, where generalized Macdonald polynomials will
depend on an additional parameter which will enter the definition of
the topological locus.

\subsection{Generalized Macdonald polynomials factorization on the
  general topological locus}
\label{sec:fact-gener-topol}
In this section we give general factorization formulas for generalized
Macdonald polynomials.  Concretely, we have found a generalization of
the factorization formula for generalized Macdonald polynomials
conjectured in~\cite{KM} to a wider topological locus. The identity
reads\footnote{Similar identity for $A=Q$ has already appeared
  in~\cite{Z} (see Eq.~(24) there). There was a minor typo in~\cite{Z}
  Eq.~(24): Kronecker symbol $\delta_{Y_2 \varnothing}$ was missing in
  the r.h.s.}:
\begin{framed}
  \begin{equation}
  \label{eq:1}
  M_{Y_1Y_2}^{(q,t)}\left( Q \Bigg| \frac{1 - A^n}{1-t^n}, \frac{
      \left( \frac{t}{q} \right)^n - 1}{1 - t^n} \right)=(-1)^{|Y_2|} t^{\frac{||Y_1^{\mathrm{T}}||^2-|Y_1|}{2}} q^{\frac{||Y_2||^2-|Y_2|}{2}}\frac{ G_{Y_1 \varnothing}^{(q,t)} (A)
    G_{Y_2 \varnothing}^{(q,t)} (AQ^{-1}) G_{ \varnothing Y_1}^{(q,t)}
    (Q^{-1}) G_{ \varnothing Y_2}^{(q,t)} (1) }{C_{Y_1}'(q,t)
    C_{Y_2}'(q,t) G_{Y_2 Y_1}^{(q,t)}(Q^{-1}) 
  }
\end{equation}
\end{framed}
where
\begin{gather}
  \label{eq:8}
  G^{(q,t)}_{AB} (x)= \prod_{(i, j) \in A} \left( 1 - q^x q^{A_i - j}
    t^{B^{\mathrm{T}}_j - i + 1} \right) \prod_{(i,j) \in B}\left(1 -
    q^x q^{-B_i + j - 1} t^{-A^{\mathrm{T}}_j + i} \right) \\
  C_Y'(q,t) = \prod_{(i,j) \in Y} (1 - q^{\mathrm{Arm}_Y(i,j)}
  t^{\mathrm{Leg}_Y(i,j)+1} ), \qquad |Y| = \sum_{i=1}^{l(Y)} Y_i,\qquad
  ||Y||^2 = \sum_{i=1}^{l(Y)} Y_i^2
\end{gather}
The original formula which appeared in~\cite{KM} is given by
\begin{framed}
  \begin{equation}
  \label{eq:2}
  M_{Y_1Y_2}^{(q,t)}\left( Q \Bigg| - \frac{1}{1-t^n}, 0
  \right)=(-1)^{|Y_1|}q^{\frac{||Y_1||^2-|Y_1|}{2}} Q^{-|Y_2|} q^{||Y_2||^2 - |Y_2|}
    t^{-\frac{||Y_2^{\mathrm{T}}||^2 - |Y_2|}{2}}\frac{ G_{ \varnothing Y_1}^{(q,t)} (Q^{-1}) G_{ \varnothing
      Y_2}^{(q,t)} (1) }{C_{Y_1}'(q,t)
    C_{Y_2}'(q,t) G_{Y_2 Y_1}^{(q,t)}(Q^{-1}) % \prod_{(i,j) \in
    % Y_1}(-1) q^{1-j} (1 - q^{\mathrm{Arm}_{Y_1}(i,j)}
    % t^{\mathrm{Leg}_{Y_1}(i,j)+1} ) \prod_{(i,j) \in Y_2}Q
    % q^{2(1-j)} t^{i-1}
    % (1 - q^{\mathrm{Arm}_{Y_2}(i,j)} t^{\mathrm{Leg}_{Y_2}(i,j)+1} )
  }
\end{equation}
\end{framed}
\noindent
It is obtained from Eq.~\eqref{eq:1} in the limit $A\to \infty$ (one
should divide both sides by $A^{|Y_1|+|Y_2|}$ to get a finite answer).

For completeness let us also give the factorization formula where the
second argument of the generalized Macdonald polynomial is nontrivial:
\begin{framed}
  \begin{equation}
    M_{Y_1Y_2}^{(q,t)}\left( Q \Bigg| 0, - \frac{1 -
        \left( \frac{t}{q} B \right)^n}{1 - t^n} \right) % = (-1)^{|Y_2|} t^{\frac{||Y_1^{\mathrm{T}}||^2-|Y_1|}{2}} q^{\frac{||Y_2||^2-|Y_2|}{2}} \frac{G_{Y_1 \varnothing}^{(q,t)} (1) G_{Y_2 \varnothing}^{(q,t)}
      % (Q^{-1}) G_{ \varnothing Y_1}^{(q,t)} (B Q^{-1}) G_{ \varnothing
      %   Y_2}^{(q,t)} (B) }{C_{Y_1}'(q,t)
      % C_{Y_2}'(q,t) G_{Y_2 Y_1}^{(q,t)}(Q^{-1})}
    = \delta_{Y_1
      \varnothing} (-1)^{|Y_2|}  q^{\frac{||Y_2||^2-|Y_2|}{2}} \frac{ G_{ \varnothing
        Y_2}^{(q,t)} (B) }{
      C_{Y_2}'(q,t)  }
  \label{eq:3}
\end{equation}
\end{framed}
\noindent
Notice the asymmetry between the two arguments of the generalized
Macdonald polynomials: while the first factorization
formula~\eqref{eq:1} has nontrivial dependence on both Young diagrams,
the second one~\eqref{eq:3} actually reduces to the
formula~\eqref{eq:33} for the ordinary Macdonald polynomials. This
peculiar feature can be traced back to the nontrivial choice of
coproduct in the DIM algebra~\cite{DIM-RTT}.

From the form of Eqs.~(\ref{eq:1}) and~(\ref{eq:3}) one could have
suspected that there is a general two-parametric factorization formula
involving both $A$ and $B$. However, it turns out that this is not the
case, as we explain below.

\subsection{Factorization identities from matrix model averages}
\label{sec:fact-ident-from}
Before presenting more generalizations of the factorization formulas
for generalized polynomials let us give here a short and simple proof
for the factorization identities obtained so far. To this end we will
employ the \textit{integral factorization identities} discovered
in~\cite{Z},~\cite{MZ}.

These identities give explicit factorized answers for $q$-Selberg
averages of generalized Macdonald polynomials. The Selberg average of
a symmetric function $f(x_i)$ is given by the following matrix
integral:
\begin{equation}
  \label{eq:13}
  \langle f(x_i) \rangle_{u,v,N,q,t}
  \stackrel{\mathrm{def}}{=} \frac{\int_0^1 d^N_qx \, \mu(u,v,N,q,t|x_i) f(x_i)}{\int_0^1 d^N_qx\, \mu(u,v,N,q,t|x_i)}
\end{equation}
where the integration measure is
\begin{equation}
  \label{eq:28}
    \mu(u,v,N,q,t|x_i) = \prod_{i \neq j} \prod_{k\geq 0} \frac{\left(
      \frac{x_i}{x_j}; q\right)_{\infty}}{\left(
      t \frac{x_i}{x_j}; q\right)_{\infty}}  \prod_{i=1}^N \left(x_i^u
    \prod_{k\geq 0} \frac{\left( x_i;q \right)_{\infty}}{\left( q^v
        x_i;q \right)_{\infty}} \right)
\end{equation}
and the Jackson $q$-integral is defined as
\begin{equation}
  \label{eq:32}
  \int_0^a d_qx g(x) = (1-q) a \sum_{n \geq 0} q^n g(q^na)
\end{equation}
One example of the factorized identity for the average considered
in~\cite{MZ} is
\begin{multline}
  \label{eq:34}
    \left \langle M_{AB}^{(q,t)}\left( q^{-u-1} t \left| p_{-n} +
        \frac{(t/q)^n - q^{nv}}{1-t^n} , -p_{-n} - \left( \frac{t}{q}
        \right)^n \frac{1 - (t/q)^n}{1-t^n} \right. \right) \right
  \rangle_{u,v,N,q,t} =\\
  =(-1)^{|A|} q^{-2|B|+u|A|} t^{|B|-|A|} t^{\sum_{(i,j)\in B} i + 2
    \sum_{(i,j)\in A} i} q^{- \sum_{(i,j)\in A} j} \times\\
  \times \frac{G_{A\varnothing}\left( t^{-N} q^{-u} \right) G_{A\varnothing}\left( t^{N-1}
      q^{v+1} \right) G_{B\varnothing}\left( t^{-N-1} q \right)
    G_{B\varnothing}\left( t^{N-2} q^{u+v+2} \right)}{ C'_A(q,t) C'_B(q,t)
    G_{BA}^{(q,t)} \left( q^{u+1} t^{-1} \right)}.
\end{multline}
Notice that the parameters of the measure also enter the arguments of
the generalized polynomials under the average sign. Let us make a
peculiar specialization of Eq.~\eqref{eq:34} and take $N=0$. What does
it mean to have \emph{zero} number of integrations? There is of course
no general answer, but for Selberg averages the definition we consider
seems very natural and can be obtained from analytic continuation in
$N$. The generalized polynomial under the average is written in terms
of power sums $p_n = \sum_{i=1}^N x_i^n$ of integration variables. For
$N=0$ power sums contain zero terms and therefore should vanish. It is
also evident that for $N=0$ there are no integrations neither in the
numerator, nor in the denominator in the definition of the average and
the integration measure is absent. Thus the l.h.s.\ of
Eq.~\eqref{eq:34} reduces to the generalized Macdonald polynomial
\emph{evaluated} at the point $p_n=0$, while the r.h.s.\ gives the
correct factorized answer, coinciding with Eq.~\eqref{eq:1}.

Notice that the topological locus parametrized by $u$ and $v$ in
Eq.~\eqref{eq:34} and by $Q$ and $A$ in Eq.~\eqref{eq:1} is
two-dimensional. This will always be the case in our considerations
since the original integral depends on three parameters, $u$, $v$, and
$N$ and we have to put $N$ to zero.

More identities can be obtained by using the symmetry of the Selberg
measure $\mu(u,v,N,q,t|x_i)$ under the change of parameters:
\begin{equation}
  \left(
    \begin{array}{c}
      u\\
      v\\
      N      
    \end{array}
\right) \to \left(
    \begin{array}{c}
      \widetilde{u}\\
      \widetilde{v}\\
      \widetilde{N}      
    \end{array}
\right) = \left(
    \begin{array}{c}
      u\\
      -v - 2 +2\beta\\
      N + \frac{v + 1 - \beta}{\beta}      
    \end{array}
\right)\label{eq:9}
\end{equation}
Since $\mu(u,v,N,q,t|x_i) =
\mu(\widetilde{u},\widetilde{v},\widetilde{N},q,t|x_i)$, the average
of any function $f(x)$ remains the same,
\begin{equation}
  \label{eq:11}
  \left\langle f(x) \right\rangle_{u,v,N,q,t} = \left\langle f(x) \right\rangle_{\widetilde{u},\widetilde{v},\widetilde{N},q,t}.
\end{equation}
Of course, if the function $f$ itself depends on the parameters $u$,
$v$ or $N$ one has to replace them with $\widetilde{u}$,
$\widetilde{v}$ or $\widetilde{N}$ respectively to get the same
average. Making the change of variables~\eqref{eq:9} in
Eq.~\eqref{eq:34} and setting $\widetilde{N}=0$ we get the
identity~\eqref{eq:3}.

Summarizing, the factorization identities for generalized Macdonald
polynomials~\eqref{eq:1},~\eqref{eq:3} follow from the integral
identity~\eqref{eq:34} in the limit $N=0$. In the next section we will
give more factorization identities involving skew generalized
Macdonald polynomials. They are proven using a similar argument.

\subsection{New formulas for skew generalized Macdonald polynomials}
\label{sec:fact-skew-gener}
We can also take the specialization $N=0$ in more general integral
factorization formulas from~\cite{MZ} (see Eqs.~(93),~(94) there). The
identities we obtain in this way involve two \emph{skew} generalized
Macdonald polynomials.  Skew generalized Macdonald polynomials are
defined similarly to the usual skew Macdonald polynomials:
\begin{equation}
  \label{eq:6}
  M_{Y_1 Y_2/Z_1 Z_2}^{(q,t)}(Q|p_n, \bar{p}_n) =
  M_{Z_1}^{(q,t)}\left( n \frac{1-q^n}{1-t^n} \frac{\partial}{\partial
      p_n} \right) M_{Z_2}^{(q,t)}\left( n \frac{1-q^n}{1-t^n}
    \frac{\partial}{\partial \bar{p}_n} \right) M_{Y_1 Y_2}^{(q,t)}(Q|p_n, \bar{p}_n)
\end{equation}
where $M_Z^{(q,t)}$ are ordinary Macdonald polynomials. Without giving
too much technical details let us write down the final results:
\begin{framed}
  \begin{multline}
  \label{eq:4}
  \sum_{Z_1, Z_2} \frac{\left( \frac{t}{q}
    \right)^{|Z_1|+|Z_2|}}{||M_{Z_1}||^2 ||M_{Z_2}||^2} M_{Y_1Y_2/Z_1
    Z_2}^{*(q,t)} \left( Q \Bigg| -\frac{1 - \left( \frac{t}{q}
      \right)^n}{1 - t^n} , - \frac{1 - B^{-n}}{1 - t^n} \right)
  M_{W_1 W_2/Z_1 Z_2}^{(q,t)}
  \left( \frac{t}{q} B Q \Bigg| 0, \frac{1 - \left( \frac{t}{q} B \right)^n}{1 - t^{-n}} \right) =\\
  = \left(\frac{t}{q B} \right)^{|Y_1|}
  t^{\frac{||Y_1^{\mathrm{T}}||^2 - |Y_1|}{2}} \left( - \frac{t Q}{q
      B}\right)^{|Y_2|} q^{-\frac{||Y_2||^2-|Y_2|}{2}}
  t^{||Y_2^{\mathrm{T}}||^2-|Y_2|} \left( -t \right)^{|W_1|}
  q^{\frac{||W_1||^2-|W_1|}{2}}\times\\
  \times  \left( \frac{q}{BQ} \right)^{|W_2|}
  q^{||W_2||^2 - |W_2|}
  t^{-\frac{||W_2^{\mathrm{T}}||^2-|W_2|}{2}}\left(
    C_{Y_1}'(q,t)C_{Y_2}'(q,t)C_{W_1}'(q,t)C'_{W_2}(q,t) \right)^{-1} \times\\
  \times \frac{z_{\mathrm{bifund}}^{\vec{Y}, \vec{W}}\left( Q^{\frac{1}{2}} ,
      \left( \frac{t}{q} B Q \right)^{\frac{1}{2}} ,  \left(
        \frac{q}{t} B \right)^{-\frac{1}{2}}
    \right)}{
    G_{Y_1 Y_2}^{(q,t)}(Q) G_{W_2 W_1}^{(q,t)}\left( \left(
        \frac{t}{q} BQ \right)^{-1}\right)}
\end{multline}
\end{framed}
\noindent
where the conjugate generalized polynomial is defined as
\begin{equation}
  \label{eq:5}
  M_{Y_1Y_2}^{*(q,t)}(Q|p_n, \bar{p}_n) =
  M_{Y_2Y_1}^{(q,t)}\left(Q^{-1}\Big|\bar{p}_n, p_n - \left( 1 -
      \left( \frac{t}{q} \right)^n \right) \bar{p}_n\right),
\end{equation}
and the norm of Macdonald polynomial is given by an explicit
expression
\begin{equation}
  \label{eq:18}
  ||M_Y||^2 = \frac{C'_Y(q,t)}{C_Y(q,t)}\qquad   C_Y(q,t) = \prod_{(i,j) \in Y} (1 - q^{\mathrm{Arm}_Y(i,j)+1}
  t^{\mathrm{Leg}_Y(i,j)} ) 
\end{equation}
The bifundamental Nekrasov function is given by
\begin{equation}
  \label{eq:10}
    z_{\mathrm{bifund}}^{\vec{Y} , \vec{W}}( Q, P, M) = G_{Y_1
    W_1}^{(q,t)} \left( \frac{Q}{M P} \right) G_{Y_1
    W_2}^{(q,t)} \left( \frac{QP}{M} \right) G_{Y_2
    W_1}^{(q,t)} \left( \frac{1}{M QP} \right) G_{Y_2
    W_2}^{(q,t)} \left( \frac{1}{M Q P} \right)
\end{equation}
There is one more identity similar to Eq.~\eqref{eq:4}:
\begin{framed}
  \begin{multline}
  \label{eq:7}
  \sum_{Z_1, Z_2} \frac{\left( \frac{t}{q}
    \right)^{|Z_1|+|Z_2|}}{||M_{Z_1}||^2 ||M_{Z_2}||^2} M_{Y_1Y_2/Z_1
    Z_2}^{*(q,t)} \left( Q \Bigg| -\frac{\left(\frac{t}{q} \right)^n -
      A^{-n}}{1 - t^{-n}} , 0 \right) M_{W_1 W_2/Z_1 Z_2}^{(q,t)}
  \left( \frac{t}{q} A Q \Bigg| \frac{1 - A^n}{1 - t^n}, - \frac{1 - \left( \frac{t}{q} \right)^n}{1 - t^n} \right) =\\
  = \left( \frac{t^2}{q}
  \right)^{|Y_1|} t^{\frac{||Y_1^{\mathrm{T}}||^2-|Y_1|}{2}}  \left( -
    \frac{Q t^2}{q} \right)^{|Y_2|} q^{-\frac{||Y_2||^2 - |Y_2|}{2}}
  t^{||Y_2^{\mathrm{T}}||^2 - |Y_2|}\left( - A \right)^{|W_1|}
  q^{\frac{||W_1||^2 - |W_1|}{2}} \times\\
  \times  \left( \frac{q}{t Q} \right)^{|W_2|}
  q^{||W_2||^2 - |W_2|} t^{-\frac{||W_2^{\mathrm{T}}||^2 - |W_2|}{2}}\left(C_{Y_1}'(q,t) C_{Y_2}'(q,t) C_{W_1}'(q,t) C_{W_2}'(q,t)\right)^{-1} \times\\
  \times \frac{z_{\mathrm{bifund}}^{\vec{Y}, \vec{W}}\left( 
      Q^{\frac{1}{2}}, \left( \frac{t}{q} A Q \right)^{\frac{1}{2}} ,
      \left( \frac{t}{q} A \right)^{\frac{1}{2}}
    \right)}{G^{(q,t)}_{Y_1 Y_2}(Q) G^{(q,t)}_{W_2 W_1} \left(
      \frac{q}{t A Q} \right)}
\end{multline}
\end{framed}
Identities~\eqref{eq:4},~\eqref{eq:7} are more general than
Eqs.~\eqref{eq:1},~\eqref{eq:3} and reduce to them in special cases.
For $Y_{1,2} = \varnothing$ Eq.~\eqref{eq:4} reduces to
Eq.~\eqref{eq:3} and for $W_{1,2} = \varnothing$ it reduces to
Eq.~\eqref{eq:1}. In Eq.~\eqref{eq:7} the situation is reversed, i.e.\
for $Y_{1,2} = \varnothing$ it reduces to Eq.~\eqref{eq:1} and for
$W_{1,2} = \varnothing$ it reduces to Eq.~\eqref{eq:3}.

\subsection{Gluing, traces and factorization of instanton sums}
\label{sec:glue-fact-inst}
The new identity~\eqref{eq:4} allows one to \emph{glue} several
factorized expressions together and then use Cauchy completeness to
obtain a factorized answer for the full sum of factorized terms. As a
simplest example we can take the \emph{trace} over Young diagrams
$\vec{Y}=\vec{W}$ in the identity~\eqref{eq:4}. In the language of
gauge theory this corresponds to making a circular quiver representing
a $U(2)$ adjoint theory, while for topological strings this gives the
partial compactification of the base of the toric fibration. In each
case, to get a meaningful result we have to set spectral parameters of
the generalized Macdonald polynomials equal to each other. For
Eq.~\eqref{eq:4} this means taking $B = \frac{q}{t}$. Thus, we set $B
= \frac{q}{t}$, $Y_1 = W_1$ and $Y_2 = W_2$ in Eq.~\eqref{eq:4} and
take the sum over Young diagrams $Y_{1,2}$ with weight
$\frac{\Lambda^{|Y_1|+|Y_2|}}{||M_{Y_1}||^2 ||M_{Y_2}||^2}$. The
r.h.s.\ of Eq.~\eqref{eq:4} then takes the form of Nekrasov instanton
partitions function for a particular value of the adjoint
hypermultiplet mass:
\begin{multline}
  \label{eq:16}
  \sum_{Y_1, Y_2} \frac{\Lambda^{|Y_1|+|Y_2|}}{||M_{Y_1}||^2
    ||M_{Y_2}||^2}\sum_{Z_1, Z_2} \frac{\left( \frac{t}{q}
    \right)^{|Z_1|+|Z_2|}}{||M_{Z_1}||^2 ||M_{Z_2}||^2} M_{Y_1Y_2/Z_1
    Z_2}^{*(q,t)} \left( Q \Bigg| -\frac{1 - \left( \frac{t}{q}
      \right)^n}{1 - t^n} , - \frac{1 - \left( \frac{t}{q}
      \right)^n}{1 - t^n} \right) M_{Y_1 Y_2/Z_1 Z_2}^{(q,t)}
  \left(  Q | 0, 0 \right) =\\
  =\sum_{\vec{Y}} \left(\frac{t^3}{q^3} \Lambda  \right)^{|\vec{Y}|}  \frac{z_{\mathrm{bifund}}^{\vec{Y}, \vec{Y}}\left(
      Q^{\frac{1}{2}} , Q^{\frac{1}{2}} , \frac{t}{q} \right)}{
    z_{\mathrm{vect}}^{\vec{Y}}\left( Q^{\frac{1}{2}} \right)}
\end{multline}

\begin{figure}[H]
  \centering
  $\parbox{4cm}{\includegraphics[width=4cm]{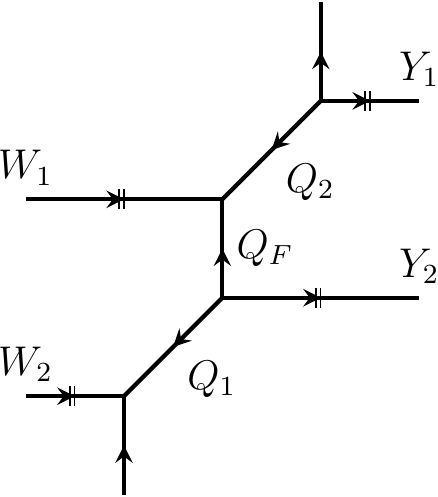}}
  \quad  \sim \quad \left\langle \sum_{EF} M_{Y_1 Y_2/EF}^{*} M_{W_1 W_2/EF}
    \right\rangle_{u,v,N}$
    \caption{Refined topological string amplitude giving the
      $q$-Selberg average of two skew generalized Macdonald
      polynomials. The number of integrations $N$ and the parameters
      of the integral $u,v$ are expressed through the K\"ahler
      parameters $Q_{1,2}$ and $Q_F$ according to Eq.~\eqref{eq:15}.}
  \label{fig:1}
\end{figure}

Now we notice that the l.h.s.\ of Eq.~\eqref{eq:16} does not depend on
the choice of basis in the space of symmetric polynomials, since it is
a trace over this space. This immediately implies that the l.h.s.\ is
in fact independent of $Q$. Choosing the basis of \emph{ordinary}
Macdonald polynomials we find that the sum factorizes into a product
of two identical sums:
\begin{multline}
  \label{eq:19}
   \sum_{Y_1, Y_2} \frac{\Lambda^{|Y_1|+|Y_2|}}{||M_{Y_1}||^2
    ||M_{Y_2}||^2}\sum_{Z_1, Z_2} \frac{\left( \frac{t}{q}
    \right)^{|Z_1|+|Z_2|}}{||M_{Z_1}||^2 ||M_{Z_2}||^2} M_{Y_1Y_2/Z_1
    Z_2}^{*(q,t)} \left( Q \Bigg| -\frac{1 - \left( \frac{t}{q}
      \right)^n}{1 - t^n} , - \frac{1 - \left( \frac{t}{q}
      \right)^n}{1 - t^n} \right) M_{Y_1 Y_2/Z_1 Z_2}^{(q,t)}
  \left(  Q | 0, 0 \right) =\\
  = \left[\sum_{Y} \frac{\Lambda^{|Y|}}{||M_Y||^2
    }\sum_Z \frac{\left( \frac{t}{q}
    \right)^{|Z|}}{||M_Z||^2} M_{Y/Z
    }^{(q,t)}  \left( -\frac{1 - \left( \frac{t}{q}
      \right)^n}{1 - t^n} \right) M_{Y/Z}^{(q,t)}
  \left(  0 \right) \right]^2
\end{multline}
One can immediately notice that
\begin{equation}
  \label{eq:20}
  M_{Y/Z}^{(q,t)}(0) = \delta_{Y Z}
||M_Y||^2, \qquad \text{and} \qquad M_{Y/Y}^{(q,t)}(p_n) = ||M_Y||^2
\end{equation}
so that the double sum in the r.h.s.\ turns into a single one:
\begin{equation}
  \label{eq:21}
  \sum_{Y} \frac{\Lambda^{|Y|}}{||M_Y||^2
    }\sum_Z \frac{\left( \frac{t}{q}
    \right)^{|Z|}}{||M_Z||^2} M_{Y/Z
    }^{(q,t)}  \left( -\frac{1 - \left( \frac{t}{q}
      \right)^n}{1 - t^n} \right) M_{Y/Z}^{(q,t)}
  \left(  0 \right) = \sum_Y \left( \frac{t}{q} \Lambda \right)^{|Y|}
  = \prod_{k \geq 1} \frac{1}{1 - \left( \frac{t}{q} \Lambda \right)^k}
\end{equation}
Eventually, we get the factorized answer for Nekrasov instanton
partition function:
\begin{equation}
  \label{eq:22}
\boxed{  \sum_{\vec{Y}} \left(\frac{t^3}{q^3} \Lambda  \right)^{|\vec{Y}|}  \frac{z_{\mathrm{bifund}}^{\vec{Y}, \vec{Y}}\left(
      Q^{\frac{1}{2}} , Q^{\frac{1}{2}} , \frac{t}{q} \right)}{
    z_{\mathrm{vect}}^{\vec{Y}}\left( Q^{\frac{1}{2}} \right)} =
  \prod_{k \geq 1} \frac{1}{\left(1 - \left( \frac{t}{q} \Lambda \right)^k \right)^2}}
\end{equation}
This is in fact nothing but the partition function of the
corresponding $2d$ CFT, which contains two bosonic field, hence power
two in the r.h.s. The example we have described is of course a trivial
one, since each term in the l.h.s.\ simplifies due to the identity
\begin{equation}
  \label{eq:23}
  z_{\mathrm{bifund}}^{\vec{Y}, \vec{Y}}\left(
    Q^{\frac{1}{2}} , Q^{\frac{1}{2}} , \frac{t}{q} \right)=
  \left( \frac{q}{t} \right)^{2|\vec{Y}|} z_{\mathrm{vect}}^{\vec{Y}}\left( Q^{\frac{1}{2}} \right)
\end{equation}
However, gluing \emph{two or more} bifundamental contributions from
Eqs.~\eqref{eq:4},~\eqref{eq:7} together one gets \emph{nontrivial}
factorization identities for linear quiver gauge theories. Also one
can take the trace to obtain circular quivers with several nodes.

\begin{figure}[h]
  \centering
  \parbox{4cm}{\includegraphics[width=4cm]{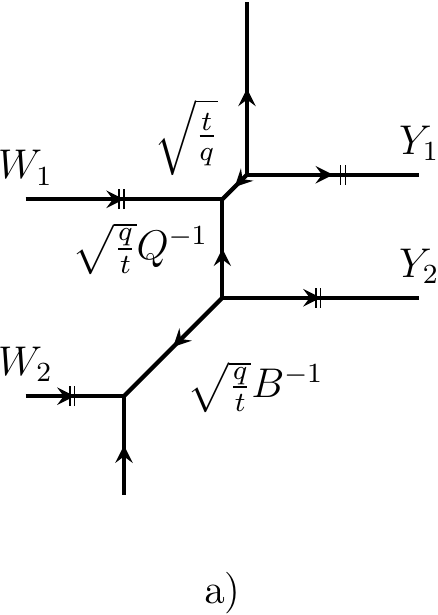}}
  \quad = \quad \eqref{eq:4}
  \hspace{2cm} \parbox{4cm}{\includegraphics[width=4cm]{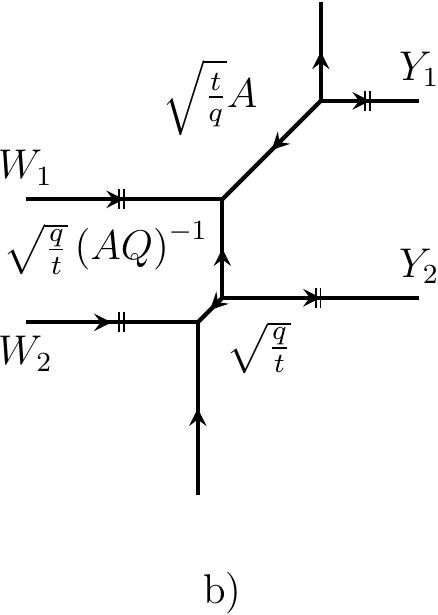}}
  \quad = \quad \eqref{eq:7}
 \caption{a) Setting $Q_1 = \sqrt{\frac{t}{q}}$ in the CY corresponds
   to setting $N=0$ in the Selberg average. For $Q_2 =
     \sqrt{\frac{q}{t} } B^{-1}$ and $Q_F = \sqrt{\frac{q}{t}}  Q^{-1}$ one arrives precisely at the factorization
   formula~\eqref{eq:4}. The lower resolved conifold piece with
   K\"ahler parameter $\sqrt{\frac{q}{t}}B^{-1}$ can be transformed by
   geometric transition into geometry containing a stack of $M$ toric
   branes, where $t^M = \frac{q}{t}B^{-1} $. b) Setting $Q_2 =
   \sqrt{\frac{q}{t}}$ in the CY corresponds to setting $\beta N= -v$
   in the Selberg average. This average is related by the
   symmetry~\eqref{eq:9} to the integral with $N=0$ number of
   integrations shown in a). For $Q_1 = \sqrt{\frac{t}{q} } A$ and
   $Q_F = \sqrt{\frac{q}{t}} \left(  A Q \right)^{-1}$ one arrives at the
   factorization formula~\eqref{eq:7}. This amplitudes corresponds to
   a stack of $M'$ toric branes on the upper horizontal leg with $M'$
   given by $t^{M'} = A$.}
  \label{fig:2}
\end{figure}

\section{Toric brane on the intermediate leg and surface operators}
\label{sec:expl-from-refin}
In this section we will demonstrate that factorization identities we
have obtained can be thought of as the amplitudes of refined
topological strings in the presence of a stack of toric branes. We
will also comment on their relation with surface operators in gauge
theory and degenerate vertex operators in $2d$ CFT.

\subsection{Refined topological amplitudes with branes}
\label{sec:refin-topol-ampl}
Selberg averages such as Eq.~\eqref{eq:34}, which we have used in our
proof of factorization in sec.~\ref{sec:fact-ident-from}, can be
identified with refined topological string amplitudes on toric CY
depicted in Fig.~\ref{fig:1}. K\"ahler parameters $Q_{1,2}$ and $Q_F$
of the CY are related to the matrix integral parameters $u$, $v$, $N$
as follows:
\begin{equation}
  \label{eq:15}
  Q_1 = t^{\frac{1}{2} - N} q^{\frac{1}{2} - v}, \qquad Q_2 =
  t^{\frac{1}{2} - N} q^{-\frac{1}{2}}, \qquad Q_F = q^{u+v+
    \frac{3}{2}} t^{N - \frac{3}{2}}
\end{equation}

In Eq.~\eqref{eq:15} one can also use $\widetilde{u}$, $\widetilde{v}$
and $\widetilde{N}$ obtained by the change of variables~\eqref{eq:9}
instead of $u$, $v$ and $N$ to get the second possible identification
between the parameters.
\begin{figure}[H]
  \centering
  \includegraphics[width=10cm]{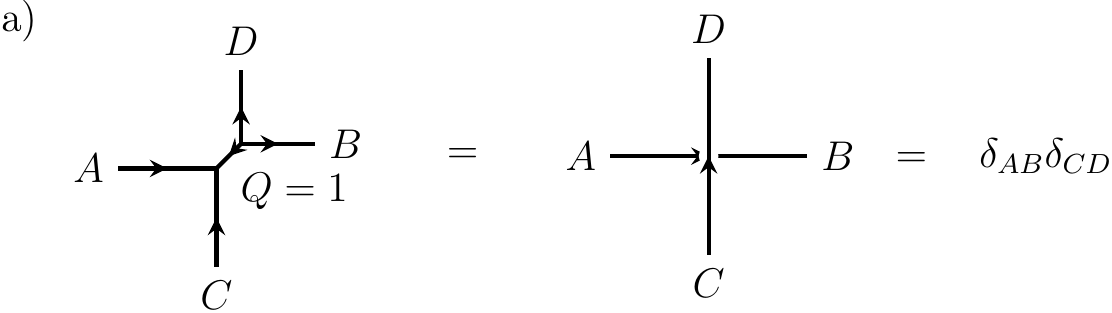}\\
    \includegraphics[width=10cm]{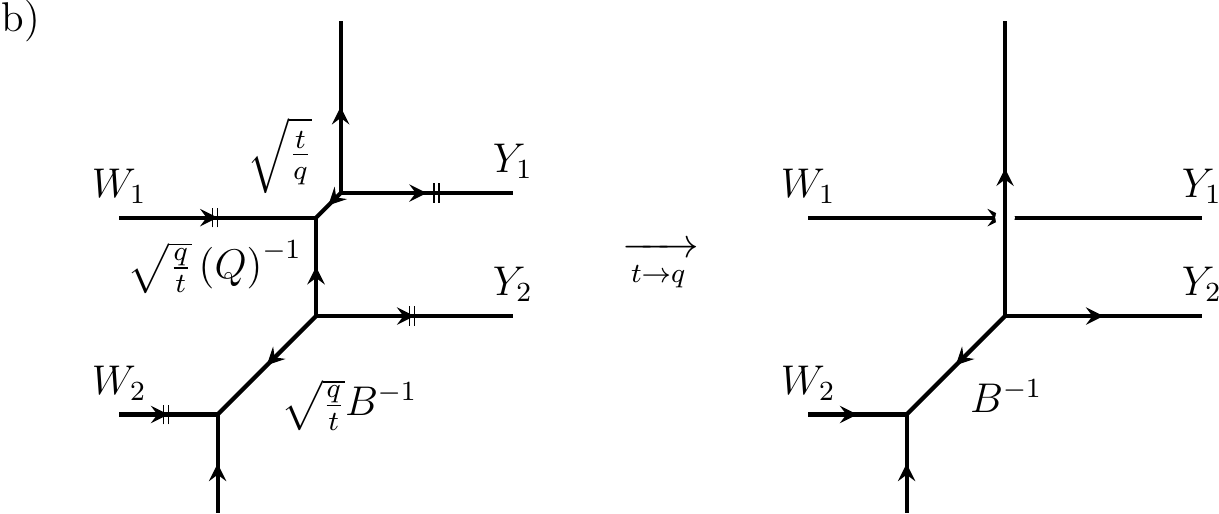}

  \caption{a) Unrefined amplitude on the resolved conifold in the
    degenerate limit $Q \to 1$ factorizes into a product of two
    separate non-interacting lines. Notice that there is no preferred
    direction in the unrefined case. b) Though the values of the
    polynomials on the topological locus are factorized into a product
    of monomials, they do not factorize into a product of independent
    terms corresponding to two horizontal lines. This happens only in
    the unrefined limit $t \to q$.}
  \label{fig:6}
\end{figure}

Factorization happens for a special value of
the parameters corresponding to $N=0$ --- the topological locus. On
this locus one of the resolved conifold pieces in the toric diagram of
the CY \emph{degenerates,} i.e.\ its K\"ahler parameters becomes
$\sqrt{\frac{q}{t}}$ or $\sqrt{\frac{t}{q}}$ as shown in
Fig.~\ref{fig:2}.

For \emph{unrefined} amplitudes degenerate resolution factorizes into
a product of two pieces as shown in Fig.~\ref{fig:6}. Each piece is
given by an explicit factorized formula, which coincides with the
factorized answer for the polynomial. However, in the unrefined case
the answers for the amplitudes from Fig.~\ref{fig:6} b) are not very
interesting since generalized Macdonald polynomials in this case
reduce to products of Schur functions, and the factorization
identities turn into the well-known formulas for quantum
dimensions~\eqref{eq:30}. They reproduce the known amplitudes in the
presence of the stack of toric branes on the intermediate leg in the
unrefined theory.

\begin{figure}[h]
  \centering
  \includegraphics[width=7cm]{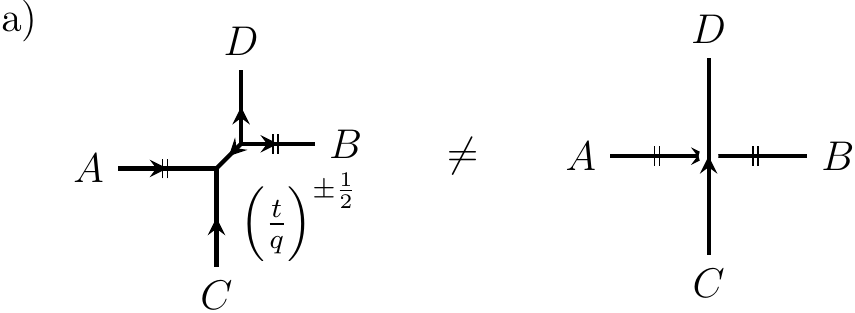}\\
    \includegraphics[width=9cm]{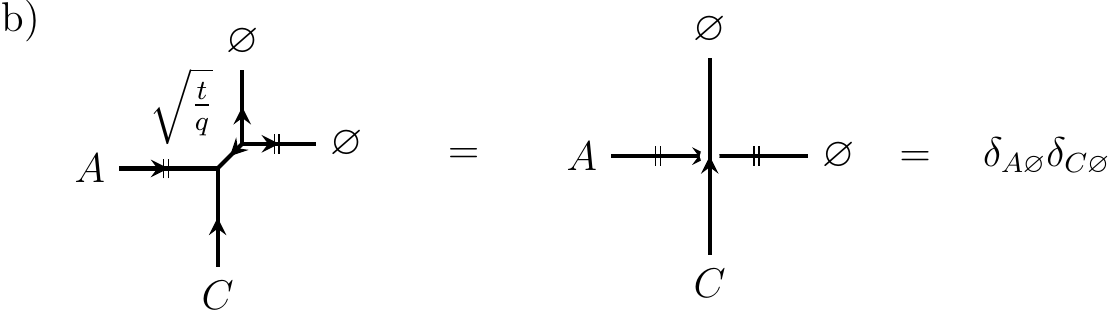}\\
  \includegraphics[width=9cm]{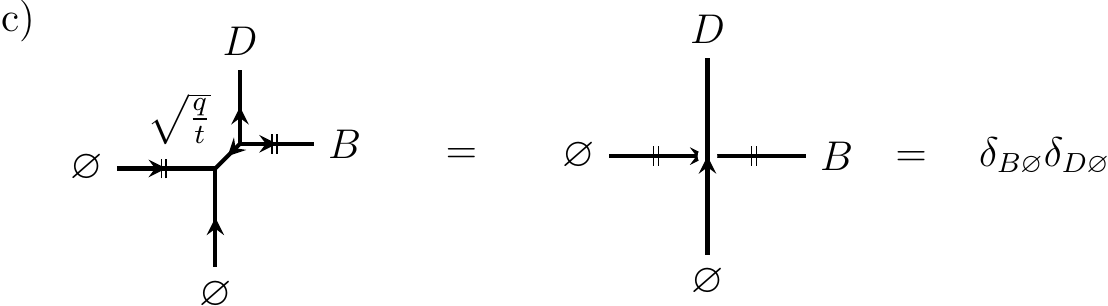}
  \caption{a) Amplitudes on degenerate resolved conifold do not
    trivialize. However, if some of the diagrams are empty, the
    amplitude reduces to the unresolved one. For two choices of
    K\"ahler parameter $\left( \frac{t}{q} \right)^{\pm \frac{1}{2}}$
    one gets different decoupling conditions, b) and c).}
  \label{fig:7}
\end{figure}

For \emph{refined} amplitudes degenerate conifold geometry does not
split into two parts. Moreover, there are \emph{two} different
degenerations of the resolved conifold with K\"ahler parameter either
$\sqrt{\frac{q}{t}}$ or $\sqrt{\frac{t}{q}}$ as shown in
Fig.~\ref{fig:7} a). The difference between these two situations is
evident from Fig.~\ref{fig:7} b) and c): when some of the legs are
empty the amplitudes do factorize and give the same result as in the
unrefined case.

The amplitudes from Fig.~\ref{fig:2} are still given by the factorized
expressions, though they cannot be separated into two noninteracting
parts as shown in Fig.~\ref{fig:6} b). We argue that this is the
natural definition of the stack of toric branes placed on the
horizontal leg of the diagrams. For a single horizontal leg the
corresponding geometry is shown in Eq.~\eqref{eq:27}. However, if we
have several horizontal legs, the vertical line will intersect several
of them. Fig.~\ref{fig:2} describes precisely this situation:
Fig.~\ref{fig:2} a) models the stack of $M$ branes on the lower
horizontal line, and Fig.~\ref{fig:2} b) represents the stack of $M'$
branes on the upper horizontal line.  The intersection of the vertical
line with the second horizontal leg is degenerate and in the unrefined
limit gives the trivial crossing from Fig.~\ref{fig:6}. In the refined
case the crossing trivializes when the corresponding diagram on the
left or right of the crossing is empty as depicted in Fig.~\ref{fig:7}
b), c).

Let us recapitulate our main point. A stack of refined toric branes
sitting on a preferred leg of the diagram interacts with all other
parallel legs. The resulting amplitude is given by the factorized
value of generalized Macdonald polynomials evaluated on the
topological locus.

\subsection{Degenerate fields and surface operators}
\label{sec:fact-cft-vert}

Topological loci can be given a natural gauge theory
interpretation. The K\"ahler moduli space of the CY is identified with
the Coulomb moduli space of the $5d$ gauge theory. The topological
locus corresponds to the root of the Higgs branch inside the Coulomb
branch. In other words, degeneration of the \emph{resolved} conifold
pieces of the toric diagram allows one to \emph{deform} the geometry
instead. This deformation corresponds to going on the Higgs
branch.

\begin{figure}[H]
  \centering
    \includegraphics[width=6cm]{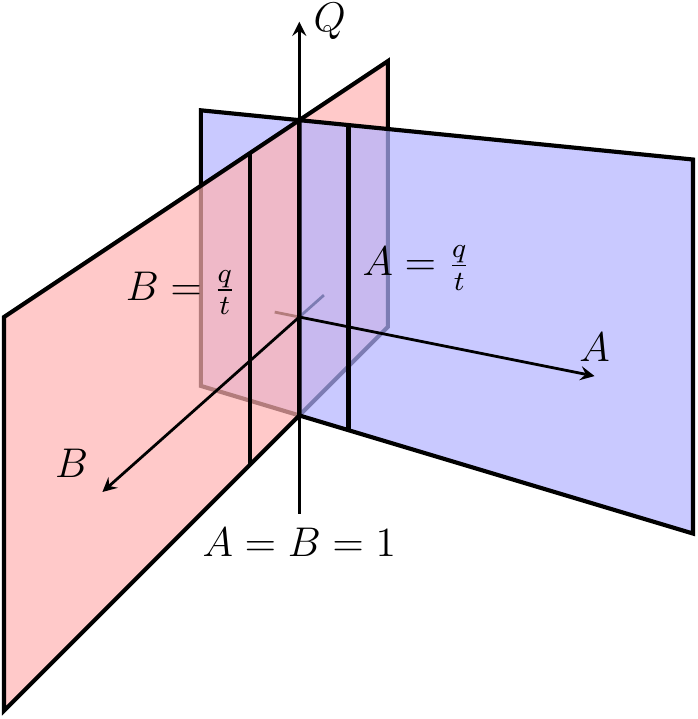}
    \caption{Schematic representation of the topological loci inside
      the K\"ahler moduli space of the strip geometry which can also
      be thought of as Coulomb moduli space of the gauge theory. The
      moduli space is parametrized by $Q$, $A$ and $B$, which are
      transformed into $Q_1$, $Q_2$ and $Q_F$ using the formulas $Q_1
      = \sqrt{\frac{t}{q}} A$, $Q_2 = \sqrt{\frac{q}{t}} B^{-1}$, $Q_F
      = \sqrt{\frac{q}{t}}( Q A)^{-1}$. There are two distinct
      topological loci, $\{ A=1\}$ and $\{ B=1\}$ (shown in red and
      blue respectively), which intersect on the line $A = B =
      1$. There are also two special lines, $A = \frac{q}{t}$ and $B =
      \frac{q}{t}$, each on its own locus, where and $Q_1 = Q_2 =
      \left( \frac{q}{t} \right)^{\pm 1}$. In the unrefined limit $t
      \to q$ the special lines coalesce with the intersection of the
      two loci.}
  \label{fig:4}
\end{figure}

The identification can also be seen directly by identifying the
parameters of the corresponding gauge theory with K\"ahler parameters
of the CY. The geometry in Fig.~\ref{fig:1} corresponds to a single
bifundamental field of mass $m$ charged under two $SU(2)$ gauge
groups, $SU(2)_L$ and $SU(2)_R$. The parameters of the theory are the
Coulomb moduli $Q_L = q^{a_L}$, $Q_R = q^{a_R}$ and the exponentiated
mass $Q_m = q^m$. They are given by the following formulas:
\begin{equation}
  \label{eq:12}
  Q_L = (Q_1 Q_F)^{\frac{1}{2}}, \qquad Q_R = \left(Q_F Q_2^{-1}\right)^{\frac{1}{2}},
  \qquad Q_m = \sqrt{\frac{t}{q}} \left(Q_1 Q_2^{-1}\right)^{\frac{1}{2}}
\end{equation}
One can immediately see that on the topological locus where we have
either $Q_L = Q_R Q_m$ or $Q_R = Q_L Q_m$. This indeed
corresponds to the origin of Higgs branches.

It is well-known that gauge theory at this point is equivalent to a
theory on a defect associated with surface operator. Factorization
formulas allow us to identify partition function of this theory with
the values of the generalized Macdonald polynomials on the topological
locus.

One final interpretation of the factorization formulas is given by
thedegenerate vertex operators in $q$-deformed $2d$ CFT. According to
the AGT relations, gauge theory we have just corresponds to vertex
operator in the Liouville theory. The Selberg integrals used in
sec.~\ref{sec:fact-ident-from} are interpreted as integrals in the DF
screening charges. Naturally, if $N=0$, the screening charges are
absent and we return to pure bosonic vertex operator
$V_{\frac{Q_1}{Q_2}}$. Matrix elements of this operator in the
generalized Macdonald basis are given by generalized Macdonald
polynomials evaluated on the topolgical locus. Schematically this can
be written as follows:
\begin{equation}
  \label{eq:14}
  \langle M^{*}_{Y_1 Y_2} (Q_R) | V_{Q_m} | M_{W_1 W_2} (Q_R Q_m)
  \rangle = \sum_{Z_1 Z_2} M^{*}_{Y_1 Y_2/Z_1 Z_2} (Q_R) M_{W_1 W_2/Z_1 Z_2} (Q_R
  Q_m) |_{\text{top locus}}
\end{equation}
As usual degenerate field obeys a difference equation. Thus,
generalized Macdonald polynomials taken on the topological locus
should also obey this equation. We hope to clarify this point in the
future.

\section{Conclusions and further prospects}
\label{sec:concl-furth-prosp}

In this paper we have presented new factorization identities for
generalized Macdonald polynomials. We proved the identities using the
technique of matrix models and related them to refined topological
string amplitudes in the presence of a stack of toric branes. We have
also identified the corresponding gauge theories and CFT vertex
operators.

It would be interesting to understand better the meaning of the
factorization identities directly in the DIM algebra. Also we would
like to investigate the difference equations satisfied by the
polynomials on the topological locus. Nekrasov-Shatashvili limit of
our construction might help to understand better the surface operators
corresponding to toric branes on the intermediate legs of the toric
diagram.

\section*{Acknowledgements}
The author thanks Y.~Kononov and N.~Sopenko for discussions. The work
of the author was supported in part by INFN, by the ERC Starting Grant
637844-HBQFTNCER and by RFBR grants 17-01-00585,
15-31-20484-mol-a-ved, 15-51-52031\_NSC, 15-51-50034\_YaF,
16-51-53034\_GFEN, 16-51-45029\_Ind.

\end{document}